\documentclass[a4paper,11pt]{article}
\usepackage{graphicx}
\begin{document}
\bibliographystyle{plain}

\def\Journal#1#2#3#4{{#1} {\bf #2}, #3 (#4)}

\def\NCA{\em Nuovo Cimento}
\def\NPA{{\em Nucl. Phys.} A}
\def\NPB{{\em Nucl. Phys.} B}
\def\PLB{{\em Phys. Lett.}  B}
\def\PRL{\em Phys. Rev. Lett.}
\def\PRC{{\em Phys. Rev.} C}
\def\PRD{{\em Phys. Rev.} D}
\def\ZPC{{\em Z. Phys.} C}

\def\ra{\rightarrow}
\def\vp{{\bf p}}
\def\al{\alpha}
\def\ab{\bar{\alpha}}

\def\be{\begin{equation}}
\def\ee{\end{equation}}
\def\bea{\begin{eqnarray}}
\def\eea{\end{eqnarray}}

\def\noin{\noindent}
\def\non{\nonumber}
\def\lab{\label}

\def\ie{{\it i.e.\ }}\def\etc{{\it etc.\ }}
\def\viz{{\it viz}}\def\eg{{\it e.g.\ }}
\def\etal{{\it et al.\ }}\def\ibid{{\it ibid\,\,\,}}

\def\bm{\boldmath}
\def\dag{\dagger}
\def\ra{\rightarrow}
\def\del{\partial}
\def\lsim{\mbox{{\scriptsize \raisebox{-.9ex}
      {$\;\stackrel{{\textstyle <}}{\sim}\,$} }} }
\def\gsim{\mbox{{\scriptsize \raisebox{-.9ex}
      {$\;\stackrel{{\textstyle >}}{\sim}\,$} }} }

\def\calL{{\cal L}}\def\calO{{\cal O}}
\def\bfq{{\bf q}}\def\bfr{{\bf r}}
\newcommand{\e}{{\mbox{e}}}
\def\CPT{{\small $\chi$PT}}
\def\cLHB{\calL^{\mbox{\scriptsize HB}}_{\rm ch}}
\def\gA{g_{\mbox{\tiny A}}}
\def\mN{m_{\mbox{\tiny N}}}

\begin{flushright}
USC(NT)-03-04
\end{flushright} 

\begin{flushleft}
{\Large
Neutrino-nucleus reactions and effective 
field theory}\footnote{Invited talk at
the First Yamada Symposium on Neutrinos 
and Dark Matter in Nuclear Physics (NDM03), Nara, Japan, 
June 2003.} 
\vspace{1cm}
\\
K. Kubodera
\\
Department of Physics and Astronomy,
University of South Carolina, Columbia,
SC 29208, USA

\end{flushleft}
\vspace{0.5cm}
\begin{abstract}

Effective field theory is believed to provide 
a useful framework for describing
low-energy nuclear phenomena 
in a model-independent fashion. 
I give here a brief account 
of the basic features of this approach,
some of its latest developments,
and examples of actual calculations carried out 
in this framework.   

\end{abstract}

\section{Introduction}

Nuclear weak-interaction processes play important roles
in many astrophysical phenomena and also 
in terrestrial experiments designed to detect
astrophysical neutrinos. 
It is obviously desirable to have reliable estimates 
of the cross sections for these processes. 
I wish to describe here some of the recent developments
in our endeavor to obtain such estimates.
I limit myself here to nuclear weak processes 
involving relatively low energy-momentum,
and, as far as calculational methods are concerned,
I'll talk about SNPA, EFT and EFT*.
These terms may seem to have popped out of alphabet soup,
but their meaning will become clear as we go along.

SNPA stands for the 
{\it standard nuclear physics approach},
which has been used extensively in describing 
a large class of nuclear properties;
a brief recapitulation of SNPA will be given later.
Recently there have been many important applications 
of {\it effective field theory} (EFT) 
to low-energy nuclear phenomena.
I would like to survey some prominent features of 
nuclear EFT.
A viewpoint that I believe worth advocating is that 
SNPA and EFT can play complementary roles.
We have recently developed a version of EFT
which allows us to take advantage of the merits 
of these two approaches.
This new method, to be referred to as EFT*, 
will also be explained in some detail below.

As concrete examples
of physical observables calculated in these methods,
I consider the following three processes:
(i) neutrino-deuteron reactions for solar neutrino
energies;
(ii) solar pp fusion;
(iii) solar Hep fusion.
I should explain why these processes are 
of particular current interest.
 
At SNO a 1-kiloton heavy water Cerenkov counter
is used to detect the solar neutrinos.
SNO can monitor the following neutrino-deuteron reactions
\be
\nu_e+d \! \rightarrow \!
 e^- \!\!+\! p \!+\! p\,,\,\,\,\,
\nu_x \!+\! d \!\rightarrow\! 
\nu_x\!+\!p \!+\! n\,,\,\,\,\,
\bar{\nu}_e \!+\! d 
\!\rightarrow\! e^+\!+\!n\!+\!n\,,\,\,\,\,
\bar{\nu}_x\!+\!d \!\rightarrow\!
\bar{\nu}_x\!+\!p\!+\!n\,,\label{nu-d}
\ee
and the pure leptonic reaction
$\nu_x+e^-\rightarrow \nu_x+e^-$.
Here $x$ stands for a neutrino 
of any flavor ($e$, $\mu$ or $\tau$).
The recent SNO experiments
\cite{ahmetal}
have established that
the total solar neutrino flux 
(summed over all flavors) agrees with the prediction
of the standard solar model~\cite{bahcall},
whereas the electron neutrino flux 
from the sun is significantly smaller 
than the total solar neutrino flux.
The amount of deficit in the electron neutrino flux
is consistent with what used to be known as
the solar neutrino problem.
These results of the SNO experiments
have given ``smoking-gun" evidence
for the transmutation of solar electron neutrinos into 
neutrinos of other flavors.
It is obvious that a precise knowledge 
of the $\nu$-$d$ reaction cross sections 
is of primary importance
in interpreting the existing and future SNO data.

The pp fusion reaction 
\be
p+p\rightarrow d+e^++\nu_e \label{eq:pp}
\ee
is the most basic solar nuclear reaction
that essentially controls the burning rate of the sun,
and hence the exact value of its cross section is a crucial
input for any further developments of solar models.
Meanwhile, the Hep fusion reaction
\be
p\,+\,^3{\rm He} \rightarrow 
\,^4{\rm He}+e^++\nu_e \label{eq:Hep}
\ee  
is important in a different context.
Although the sun very rarely uses Hep to 
produce $^4$He, Hep generates highest-energy solar neutrinos
whose spectrum extends beyond the maximum energy
of the $^8$B neutrinos. 
So, even though the flux of the Hep neutrinos 
is small, it can distort the higher end of 
the $^8$B neutrino spectrum~\cite{kuz65}, and this distortion
can affect the interpretation of the
results of a recent Super-Kamiokande experiment
\cite{SuperK-Hep}.

\section{Calculational frameworks}

\subsection{Standard nuclear physics approach (SNPA)}

As is well known, the phenomenological potential 
picture has been highly successful in describing 
many kinds of nuclear phenomena.
In this picture an A-nucleon system
is described by a Hamiltonian of the form
\be
H\,=\,\sum_i^A t_i + \sum_{i<j}^A V_{ij}
+\sum_{i<j<k}^A V_{ijk}+ \cdots\,, \label{HSNPA}
\ee 
where $t_i$ is the kinetic energy of the $i$-th nucleon,
$V_{ij}$ is a phenomenological two-body potential
between the $i$-th and $j$-th nucleons,
$V_{ijk}$ is a phenomenological three-body potential, 
and so on.
Since the interactions involving three or more nucleons 
are known to play much less important roles than
the two-body interactions, we shall be mainly 
concerned with $V_{ij}$.
Once the Hamiltonian $H$ is specified,
the nuclear wave function $|\Psi\!>$
is obtained by solving the Shr\"{o}dinger equation
\be
H|\Psi\!>\,=\,E|\Psi\!>\,.\label{Sch}
\ee
It is to be noted that the progress of 
numerical techniques for solving eq.(\ref{Sch}) has reached 
such a level \cite{cs98} that the wave functions 
of low-lying levels for light nuclei
can now be obtained practically with no approximation
(once the validity of eq.(\ref{Sch}) is accepted).
This frees us from the 
``usual" nuclear physics complications 
that arise as a result of
truncation of nuclear Hilbert space 
down to certain model space
(such as shell-model configurations, 
cluster-model trial functions, etc.)

There is large freedom in choosing
possible forms of $V_{ij}$ 
apart from a well-established requirement that,
as the inter-nucleon distance
$r_{ij}$ becomes sufficiently large,
$V_{ij}$ should approach the
one-pion exchange Yukawa potential.
For the model-dependent short-range part 
of $V_{ij}$, one needs to assume 
certain functional forms 
and fix the parameters appearing therein
by demanding that the solutions of eq.(\ref{Sch}) 
for the A=2 case reproduce the nucleon-nucleon
scattering data 
(typically up to the pion-production threshold energy)
as well as some of the deuteron properties.
There are by now several so-called 
{\it modern high-precision}
{\it phenomenological} N-N potential
that can reproduce all the existing two-nucleon data
with normalized $\chi^2$ values close to 1.
These potentials differ significantly
in the ways they parametrize 
short-range physics,
and, as a consequence,
they exhibit substantial difference
in their off-shell behavior.
To what extent this arbitrariness may affect
the observables of our concern will be discussed
below.

In normal circumstances,
nuclear responses
to external electroweak probes are 
given, to good approximation,  
by one-body terms; these are also called
the impulse approximation (IA) terms.
To obtain higher accuracy, however, 
one must also consider exchange current (EXC) 
terms,
which represent the contributions
of nuclear responses involving two or more nucleons.
In particular, if for some reason 
the IA contributions are suppressed, 
then it becomes essential to take account of 
the EXC contributions.
These exchange currents 
(usually taken to be two-body operators)
are derived from one-boson exchange diagrams,
and the vertices featuring in the relevant diagrams
are determined to satisfy the low-energy theorems
and current algebra \cite{crit}.
We refer to a formalism based on this picture
as the {\it standard nuclear physics approach}
(SNPA). (This is also called a potential model
in the literature.)
Schematically, the nuclear matrix element in SNPA
is given by 
\be
{\cal M}_{fi}^{SNPA}\,=\,
<\!\Psi_{\!f}\,|\sum_\ell^A{\cal O}_\ell
+\sum_{\ell<m}^A{\cal O}_{\ell m}
\,|\Psi_i\!>\,, \label{ME-SNPA}
\ee
where the initial (final) nuclear wave function,
$\Psi_i$ ($\Psi_f$), is a solution of eq.(\ref{Sch}),
and ${\cal O}_\ell$ and ${\cal O}_{\ell m}$ are, respectively,
the one-body and two-body transition operators for 
a given electroweak process.

SNPA has been used extensively 
for describing nuclear electroweak processes
in light nuclei, 
and general good agreement found
between theory and experiment~\cite{cs98}
gives a strong indication 
that SNPA practically captures much 
of the physics involved.

\subsection{Effective field theory (EFT)}

Although SNPA has been scoring undeniable successes
in correlating and explaining a vast variety 
of data,
it is still important from a formal point of view
to raise the following issues.
First, since hadrons and hadronic systems
(such as nuclei) are governed 
by quantum chromodynamics (QCD),
one should ultimately be able to relate SNPA with QCD,
but this relation has not been established.
In particular, while chiral symmetry is known 
to be a fundamental symmetry of QCD,
the formulation of SNPA is largely disjoint from
this symmetry.
Secondly, in SNPA, even for describing low-energy phenomena,
we start with a ``realistic" phenomenological potential
which is tailored to encode short-range (high-momentum)
and long-range (low-momentum) physics
simultaneously.  This mixing of the two different scales
seems theoretically dissatisfactory 
and can be pragmatically inconvenient. 
Thirdly, in writing down a phenomenological Lagrangian
for describing the nuclear interaction and 
nuclear responses to the electroweak currents, 
SNPA is not equipped with a clear guiding principle;  
it is not clear whether there
is any identifiable expansion parameter
that helps us to control the possible forms of terms in 
the Lagrangian and that provides a general
measure of errors in our calculation.
To address these and other related issues,
a new approach based on EFT 
was proposed~\cite{wei90}
and it has been studied with great intensity;
for reviews, see,
\cite{bkm}-\cite{br02}.
  
The general idea of EFT is in fact very simple.
In describing phenomena
characterized by a typical energy-momentum scale $Q$,
we expect 
that we need not include in our Lagrangian 
those degrees of freedom 
that pertain to energy-momentum scales 
much higher than $Q$.
This expectation motivates us to
introduce a cut-off scale $\Lambda$
that is sufficiently larger than $Q$
and we classify our fields 
(to be generically represented by $\phi$)
into two groups: high-frequency fields 
$\phi_{\mbox{\tiny H}}$
and low-frequency fields $\phi_{\mbox{\tiny L}}$.
By eliminating (or {\it integrating out})
$\phi_{\mbox{\tiny H}}$,
we arrive at an {\it effective} Lagrangian 
that only involves $\phi_{\mbox{\tiny L}}$
as explicit dynamical variables.  
Using the notion of path integrals,
the effective Lagrangian ${\cal L}_{{\rm eff}}$
is related to the original Lagrangian ${\cal L}$ as
\be
\int\![d\phi]\e^{{\rm i}\int d^4x{\cal L}(\phi)}
= \int\![d\phi_{\mbox{\tiny L}}]
\e^{{\rm i}\int d^4x{\cal L}_{\rm eff}
(\phi_{\mbox{\tiny L}})}\,.\label{EFTdef}
\ee

One can show that ${\cal L}_{{\rm eff}}$ defined 
by eq.(\ref{EFTdef})
inherits the symmetries 
(and the patterns of symmetry breaking, if there are any)
of the original Lagrangian ${\cal L}$.
It also follows that
${\cal L}_{{\rm eff}}$ should be
the sum of all possible monomials of 
$\phi_{\mbox{\tiny L}}$ and their derivatives
that are consistent with the symmetry requirements
dictated by ${\cal L}$.
Since a term involving $n$ derivatives
scales like $(Q/\Lambda)^n$,
the terms in ${\cal L}_{{\rm eff}}$ can be organized
into a perturbative series 
in which $Q/\Lambda$ serves as an expansion parameter.
The coefficients of terms in 
this expansion scheme are called
the low-energy constants (LECs).
Insofar as  all the LEC's up to a specified order $n$
can be fixed either from theory or from fitting 
to the experimental values of the relevant observables,
${\cal L}_{{\rm eff}}$ serves as 
a complete (and hence model-independent) Lagrangian
to the given order of expansion.


Having outlined the basic idea of EFT,
we now discuss specific aspects of EFT 
as applied to nuclear physics.
The underlying Lagrangian ${\cal L}$ 
in this case is the QCD Lagrangian ${\cal L}_{QCD}$,
whereas, for the typical nuclear physics
energy-momentum scale
$Q\ll \Lambda_{\chi}\sim 1$ GeV,
the effective degrees of freedom
that would feature in ${\cal L}_{{\rm eff}}$ 
are hadrons rather than the quarks and gluons.
It is a non-trivial task
to apply the formal definition in eq.(\ref{EFTdef})
to derive ${\cal L}_{{\rm eff}}$ 
written in terms of hadrons
starting from ${\cal L}_{QCD}$;
the hadrons cannot be straightforwardly
identified with the low-frequency field, 
$\phi_L$ in eq.(\ref{EFTdef}),
in the original Lagrangian.
At present, the best one could do is
to resort to symmetry considerations
and the above-mentioned expansion scheme. 
Here chiral symmetry plays an important role.  
We know that chiral symmetry is spontaneously
broken, generating the pions
as Nambu-Goldstone bosons;\footnote{
We limit ourselves here to SU(2)$\times$SU(2)
chiral symmetry.}
or chiral symmetry is realized in the Goldstone mode.
This feature can be incorporated 
by assigning suitable chiral transformation properties
to the Goldstone bosons
and writing down all possible chiral-invariant
terms up to a specified chiral order \cite{geo84}.
The above consideration presupposes 
exact chiral symmetry in ${\cal L}_{QCD}$.
In reality, ${\cal L}_{QCD}$ contains
small but finite quark mass terms,
which explicitly violate chiral symmetry
and lead to a non-vanishing value
of the pion mass $m_\pi$.
Again, there is a well-defined method 
to determine what terms are needed
in the Goldstone boson sector 
to represent the effect of 
explicit chiral symmetry breaking
\cite{geo84}.
These considerations lead to an EFT 
called chiral perturbation theory 
($\chi$PT)~\cite{wei79,gl84}.
The successes of $\chi$PT in
the meson sector are well known;
see, {\it e.g.,} \cite{bkm}. 

A problem we encounter in extending $\chi$PT
to the nucleon sector is that,
as the nucleon mass $\mN$ is comparable to 
the cut-off scale $\Lambda_{\chi}$, 
a simple application of expansion in $Q/\Lambda$
does not work.
This problem can be circumvented by employing 
heavy-baryon chiral perturbation theory (HB$\chi$PT),
which essentially consists in shifting 
the reference point of the nucleon energy 
from 0 to $\mN$ and in integrating out 
the small component of the nucleon field 
as well as the anti-nucleonic degrees of freedom.
An effective Lagrangian in HB$\chi$PT therefore involves 
as explicit degrees of freedom
the pions and the large components 
of the redefined nucleon field.
HB$\chi$PT has as expansion parameters
$Q/\Lambda_{\chi}$, $m_\pi/\Lambda_{\chi}$
and $Q/\mN$.
Since $\mN\approx \Lambda_{\chi}$,
it is convenient to combine chiral 
and heavy-baryon expansions
and introduce the chiral index ${\bar \nu}$ 
defined by $\bar{\nu}=d+(n/2)-2$.
Here $n$ is the number of fermion lines 
that participate in a given vertex,
and $d$ is the number of derivatives
(with $m_\pi$ counted as one derivative). 
A similar power counting scheme can also
be introduced
for Feynman diagrams as well.
According to Weinberg \cite{wei90},
the contribution of 
a Feynman diagram that contains $N_A$ nucleons,
$N_E$ external fields,
$L$ loops and $N_C$ disjoint parts
can be shown to scale like 
$(Q/\Lambda)^\nu$, where
the chiral index $\nu$ is defined as 
\be
\nu = 2 L + 2 (N_C-1) + 2 - (N_A+N_E) + 
\sum_i \bar \nu_i\,,\lab{eq:nu}
\ee
with the summation running over all the vertices
contained in the Feynman diagram.
HB$\chi$PT has been used with great success
to the one-nucleon sector \cite{bkm}.

However, HB$\chi$PT cannot be applied 
in a straightforward manner 
to nuclei that contain more than one nucleon.
The reason is that nuclei involve 
very low-lying excited states,
and the existence 
of this small energy scale
upsets the original counting rule~\cite{wei90}.
This is analogous to a problem
one encounters in ordinary quantum mechanics
when a system allows for low-lying intermediate states 
that spoil perturbation expansion.
Weinberg proposed to avoid this difficulty
as follows.
Classify Feynman diagrams into two groups,
irreducible and reducible diagrams.
Irreducible diagrams are those
in which every intermediate state
has at least one meson in flight;
all others are classified as reducible diagrams.
We then apply the above-mentioned chiral counting rules
only to irreducible diagrams.
The contribution of all the two-body irreducible diagrams 
(up to a specified chiral order)
is treated as an effective potential
(to be denoted by $V_{ij}^{EFT}$)
acting on nuclear wave functions.
Meanwhile, the contributions of reducible diagrams
can be incorporated~\cite{wei90} 
by solving the Schr\"odinger equation
\be
H^{EFT}|\Psi\!>^{EFT}\,=\,E|\Psi\!>^{EFT}\,,
\label{Sch-EFT}
\ee
where
\be
H^{EFT}\,=\,\sum_i^A t_i + \sum_{i<j}^A V_{ij}^{EFT}\,, 
\label{HEFT}
\ee 
We refer to this two-step procedure as
{\it nuclear} \CPT,
or, to be more specific,  
{\it nuclear} \CPT\  
in the Weinberg scheme.\footnote{
This is often called the $\Lambda$-counting scheme
\cite{lep99}.}

To apply nuclear \CPT\ to a process 
that involves (an) external current(s),
we derive a nuclear transition operator 
${\cal T}$ 
by evaluating the complete set of 
all the irreducible diagrams
(up to a given chiral order $\nu$) 
involving the relevant external current(s).
To preserve consistency in chiral counting, 
the  nuclear matrix element of ${\cal T}$ 
must be calculated with the use of nuclear 
wave functions which are governed
by nuclear interactions that represent
all the irreducible A-nucleon diagrams 
up to $\nu$-th order. 
Thus, a transition matrix in nuclear EFT
is given by
\be
{\cal M}_{fi}^{EFT}\,=\,
<\!\Psi_{\!f}^{EFT}\,|\sum_\ell^A{\cal O}_\ell^{EFT}
+\sum_{\ell<m}^A{\cal O}_{\ell m}^{EFT}
\,|\Psi_i^{EFT}\!>\,, \label{ME-EFT}
\ee
where the superscript, ``EFT", 
means that the relevant quantities are obtained
according to EFT as described above.
If this program is carried out exactly, 
it would constitute an {\it ab initio} calculation.
We note that in EFT we know exactly
at what chiral order three-body operators 
start to contribute to ${\cal T}$, 
and that, to chiral orders 
relevant to the applications
described below, there is no need for 
three-body operators.
With this understanding, 
we have retained only one- and two-body operators 
in eq.(\ref{ME-EFT}). 
This unambiguous classification of 
transition operators according to their chiral orders
is a great advantage of EFT,
which is missing in eq.(\ref{ME-SNPA}).

I should point out that there exists
an alternative form of nuclear EFT
based the power divergence subtraction (PDS) scheme. 
The PDS scheme proposed 
by Kaplan, Savage and Wise 
in their seminal papers~\cite{ksw}
uses a counting scheme (often called Q-counting)
that is different from the Weinberg scheme.
An advantage of the PDS scheme is
that it maintains formal chiral invariance,
whereas the Weinberg scheme loses
manifest chiral invariance.
In many practical applications, however,
this formal problem is not worrisome
up to the chiral order of our concern,
{\it viz.}, the chiral order up to which 
our irreducible diagrams are to be evaluated.
Although many important results have been obtained
in the PDS scheme (for a review, see \eg 
\cite{beaetal01}),
I concentrate here on the Weinberg scheme, 
as this is a framework in which
our own work has been done.

I also remark that,
if we are interested in
low-energy nuclear phenomena the typical
energy-momentum scale of which is $Q \ll m_\pi$,
even the pions may be regarded
as ``heavy" particles and can be eliminated
from ${\cal L}_{{\rm eff}}$.

\subsection{Hybrid EFT}

In the preceding subsection
we emphasized the formal merits of nuclear EFT.
In actual calculations, however, 
the following two aspects need to be considered.
First, it is still a big challenge
to generate, strictly within the EFT framework,
nuclear wave functions the accuracy of which
is comparable to that of SNPA wave functions.
Secondly, as mentioned earlier,
the chiral Lagrangian, ${\cal L}_{{\rm eff}}$,
is definite only when the values of 
all the relevant LECs are fixed,
but there may be cases where this condition
cannot be readily met.
A pragmatic solution to the first problem
is to use in eq.(\ref{ME-EFT}) wave functions
obtained in SNPA; 
we refer to this eclectic approach as hybrid EFT.
Thus a nuclear transition matrix element in hybrid EFT
is given by 
\be
{\cal M}_{fi}^{hyb-EFT}\,=\,
<\!\Psi_{\!f}^{SNPA}\,|\sum_\ell^A{\cal O}_\ell^{EFT}
+\sum_{\ell<m}^A{\cal O}_{\ell m}^{EFT}
\,|\Psi_i^{SNPA}\!>\,, \label{ME-hybrid}
\ee
Since, as mentioned, the NN interactions
that generate SNPA wave functions
reproduce accurately the entirety of
the two-nucleon data, the adoption of eq.(\ref{ME-hybrid})
is almost equivalent to using the empirical data themselves
to control the initial and final nuclear wave functions.
In the purely theoretical context of deriving
the nuclear interactions based on EFT, 
hybrid EFT may be deemed as a ``regression".
But, if our goal is to obtain 
a transition matrix element as accurately as possible
with the maximum help of available empirical input,
then hybrid EFT does have a legitimate status,
so long as the afore-mentioned off-shell problem
and the contributions of three-body (and higher-body) 
interactions are properly addressed.  These two points
will be discussed later in this talk.

The calculations reported in Refs.~\cite{pkmr99,pc00}
seem to render support for hybrid EFT.
There, the nuclear matrix elements in the A=2 systems
for one-body operators (or IA terms)
calculated with the use of EFT-generated wave functions 
were found to be very close to those calculated with 
the SNPA wave functions.
Thus EFT and hybrid EFT should give practically
the same IA matrix elements.
Meanwhile, it is generally expected that the ratio of 
the two-body EXC contributions
to those of the IA operators should 
be much less sensitive to the details 
of the nuclear wave functions
than the absolute values are.
It therefore seems reasonable
to rely on \CPT\ for deriving transition operators
and evaluate their matrix elements
using the realistic wave functions obtained in SNPA,
and in this sense hybrid EFT is more than   
a mere expedient.

The issue of possible unknown LECs
will be discussed in connection with EFT* 
in the next subsection.

\subsection{EFT*}

Hybrid EFT can be used
for complex nuclei (A = 3, 4, ...) 
with essentially the same accuracy and ease
as for the A=2 system.\footnote{
Here I am ignoring ``purely technical" complications
that can grow in actual numerical calculations 
for higher-A systems.}
We should reemphasize in this connection
that, in A-nucleon systems (A$\ge$3),
the contributions of transition operators
involving three or more nucleons
are intrinsically suppressed 
according to chiral counting, 
and hence, up to a certain chiral order,
a transition operator in an A-nucleon system
consists of the same EFT-based
1-body and 2-body terms
as used for the two-nucleon system.
Then, since SNPA provides high-quality wave functions
for the A-nucleon system,
one can calculate ${\cal M}_{fi}^{hyb-EFT}$
with precision comparable to that for 
the corresponding two-nucleon case.

Now, in most practical cases, the one-body
operator, ${\cal O}_\ell^{EFT}$, 
is free from unknown LECs.  So let us 
concentrate on the two-body operator, 
${\cal O}_{\ell m}^{EFT}$,
and suppose that ${\cal O}_{\ell m}^{EFT}$ 
under consideration contains an LEC (call it $\kappa$)
that cannot be determined with the use of 
A=2 data alone.
It is possible that 
an observable (call it $\Omega$)
in a A-body system (A$\ge$3)
is sensitive to $\kappa$ and 
that the experimental value of $\Omega$
is known with sufficient accuracy.
Then we can determine $\kappa$ 
by calculating ${\cal M}_{fi}^{hyb-EFT}$
responsible for $\Omega$
and adjusting $\kappa$ to reproduce 
the empirical value of $\Omega$.
Once $\kappa$ is fixed this way,
we can make {\it predictions} for any other 
observables for any other nuclear systems
that are controlled by the same transition
operators.
When hybrid EFT is used in this manner,
we refer to it as EFT*.\footnote{
It is also called {\it more effective} 
effective field theory 
(MEEFT)~\cite{pkmr01,PMS-pp,PMS-hep}.}

EFT* is the most efficient existing formalism
for correlating various observables in different
nuclei, 
using the transition operators
controlled by EFT.
A further notable advantage of EFT*
is that, since correlating the observables 
in neighboring nuclei is likely to serve as 
an additional renormalization,
the possible effects of higher chiral order terms
and/or off-shell ambiguities 
can be significantly suppressed
by the use of EFT*.\footnote{
EFT* should be distinguished from
an earlier naive hybrid EFT model
wherein the short-range terms were
dropped altogether using an intuitive argument
based on short-range NN repulsion.}
I will come back to this point later,
when we discuss concrete examples.

\section{Numerical results}

We now discuss the applications
of the above-described calculational methods
to the three processes of our concern:
pp fusion, Hep fusion, and the $\nu$-$d$ reaction.
A common feature of these reactions is that
a precise knowledge of the Gamow-Teller (GT) 
transition matrix elements is crucial
in estimating their cross sections.
We therefore concentrate on the GT transitions.
I will show here, following 
Refs.~\cite{PMS-pp,PMS-hep,PMS-pphep},
that the idea of EFT* can be used very nicely 
for this group of reactions.

We can argue (see, {\it e.g.,} \cite{PMS-pphep})  
that 1-body IA operators for the GT transition
can be fixed unambiguously from the available
1-body data.
As for the 2-body operators, 
to next-to-next-to-next-to-leading order (N$^3$LO)
in chiral counting,
there appears one unknown LEC that cannot be
at present determined from data for the A=2 systems.
This unknown LEC, 
denoted by $\hat{d}_R$ in \cite{pkmr98b},
parametrizes the strength
of contact-type four-nucleon coupling
to the axial current.
Park {\it et al.\ }\cite{PMS-pp,PMS-hep,PMS-pphep}
noticed that the same LEC, $\hat{d}_R$, 
also features as the only unknown parameter
in the calculation of the tritium $\beta$-decay rate
$\Gamma_{\beta}^t$,
and they proposed to use EFT*
to place a constraint on $\hat{d}_R$
from the experimental information on $\Gamma_{\beta}^t$.
Since the empirical value
of $\Gamma_{\beta}^t$ is known with high precision,
and since the accurate wave functions of 
$^3$H and $^3$He are available
from a well-developed 
variational calculation in SNPA~\cite{Metal},
we can determine $\hat{d}_R$
with sufficient accuracy for our purposes.
Once the value of $\hat{d}_R$ is determined this way, 
we can carry out parameter-free EFT* calculations 
for pp-fusion~\cite{PMS-pp,PMS-pphep},
Hep fusion~\cite{PMS-hep,PMS-pphep},
and the $\nu$-$d$ reactions~\cite{andetal-nud}.
I present here a brief summary of the results of
these calculations.  

Before doing that, we need to discuss 
the important role of momentum cutoff in EFT.
As emphasized before, the effective Lagrangian
${\cal L}_{eff}$ is, by construction, valid
only below the specified cutoff scale $\Lambda$.
Needless to say, this basic constraint should
be respected in our nuclear EFT calculations,
and for that 
we must make sure that nuclear intermediate states 
involved in the computation of eq.(\ref{ME-EFT})
do not get out of this constrained world.
It is reasonable to implement this constraint
by requiring that the two-nucleon relative momentum
should be smaller than $\Lambda$;
Park {\it et al.\ }used a Gaussian cutoff function
proportional to 
$\exp(-\vec{p}^2/\Lambda^2)$
but its detailed form should not be too important. 
As a reasonable range of the value of $\Lambda$
we may choose: 
500 MeV $\lsim \Lambda \lsim$ 800 MeV,
where the lower bound is dictated by
the requirement that $\Lambda$
should be sufficiently larger than 
the pion mass (to fully accommodate pion physics),
while the upper bound reflects the fact 
that our EFT is devoid of the $\rho$ meson.

For a given value of $\Lambda$ within the above range,
$\hat{d}_R$ is tuned to reproduce 
$\Gamma_{\beta}^t$,
and then the cross sections for pp-fusion, 
Hep fusion and the $\nu d$ reactions are calculated.
Before giving a (brief) account of 
the individual results,
I should point out a notable common feature.
Although the optimal value of $\hat{d}_R$ 
varies significantly as a function of $\Lambda$,
the observables (in our case the above three reaction
cross sections) exhibit remarkable stability
against the variation of $\Lambda$
(within the above-discussed physically reasonable range).
This stability may be taken as an indication
that the use of EFT* for
inter-correlating the observables in neighboring nuclei
{\it effectively} renormalizes various effects, 
such as the contributions of higher-chiral order terms,
mismatch between the SNPA and EFT wave functions,
etc.
This stability is essential in order 
for EFT* to maintain its predictive power. 

Park {\it et al.}~\cite{PMS-pp,PMS-pphep}
used this EFT* method
to calculate the rate of pp fusion,
$pp\!\rightarrow\!e^+\nu_e d$.
The result expressed in terms of
the threshold $S$-factor is
\be
S_{pp}(0)=3.94\!\times\!(1\pm0.005)
\times 10^{-25}\,{\rm MeV\, b}\,.\label{Spp}
\ee
It has been found that $S_{pp}(0)$ changes 
only by $\sim$0.1\%
against changes in $\Lambda$,
assuring thereby the robustness
of the prediction provided by EFT*.
The EFT* result, eq.(\ref{Spp}),
is consistent with that obtained in SNPA by Schiavilla 
{\it et al.}~\cite{schetal98}. 
Meanwhile, the fact that EFT* allows us 
to make an error estimate [as given in 
eq.(\ref{Spp})] is a notable advantage
over SNPA.
The details on how we arrive at this error estimate
can be found in \cite{PMS-pp,PMS-pphep}.
Here I just remark that the error 
indicated in eq.(\ref{Spp})
represents an improvement 
by a factor of $\sim$10
over the previous results
based on a simple hybrid EFT~\cite{pkmr98b}.

We now discuss the application of 
the same EFT* method to the
Hep fusion reaction,
$p^3{\rm He}\!\rightarrow\! 
e^+\nu_e\,^4{\rm He}$ \cite{PMS-hep,PMS-pphep}.
An accurate estimation of this cross section
is a particularly challenging task because:
(1) the contribution of the leading-order
1-body GT operator is highly suppressed
due to the approximate wave function orthogonality,
and (2) there is a strong cancellation between 
the 1-body and 2-body GT matrix elements
\cite{crsw,Metal}.
Park {\it et al.}'s EFT* 
calculation~\cite{PMS-pp,PMS-pphep}
give for the threshold S-factor 
\be
S_{Hep}(0)= (8.6\pm1.3)\times10^{-20}\,
{\rm keV\,b}\,,\label{SHep} 
\ee
where the error spans the range of 
the $\Lambda$ dependence for 
$\Lambda$ = 500-800 MeV.
Again, the EFT* result agrees with that
obtained in SNPA by Marcucci {\it et al.\ }
\cite{Metal}:
$S_{Hep}(0)= 9.64\times10^{-20}\,{\rm keV\,b}$.
The above-mentioned large cancellation between
the 1-body and 2-body contributions in this case
amplifies the cutoff dependence of  
$S_{Hep}(0)$, but the error quoted in 
eq.(\ref{SHep}) is still small enough 
for the purpose of analyzing the existing 
Super-Kamiokande data \cite{SuperK-Hep}.
 
We now move to the $\nu$-$d$ reactions, eq.(\ref{nu-d}).
It is useful to give a short but general survey 
of all the recent results obtained in SNPA, EFT
and EFT*, and that's what I am going to do here.
Within SNPA a detailed calculation of 
the $\nu$-$d$ cross sections, $\sigma(\nu d)$,
was carried out by Nakamura, Sato,
Gudkov and myself~\cite{NSGK},\footnote{
For a review of the earlier SNPA calculations,
see \cite{kn94}.}
and this calculation was recently updated
by Nakamura {\it et al.}\ (NETAL)~\cite{NETAL}.
As demonstrated in Ref.\cite{crsw},
the SNPA exchange currents for the GT transition
are dominated 
by the $\Delta$-particle excitation diagram,
and the reliability of estimation
of this diagram depends on the precision
with which the coupling constant 
$g_{\pi N\Delta}$ is known.
NETAL fixed $g_{\pi N\Delta}$
by fitting the experimental value of $\Gamma_{\beta}^t$,
the tritium $\beta$-decay rate,
and proceeded to calculate $\sigma(\nu d)$.
Meanwhile, Butler, Chen and Kong (BCK)~\cite{EFT} 
carried out
an EFT calculation of the $\nu$-$d$ cross sections,
using the PDS scheme~\cite{ksw}.
The results obtained by BCK
agree with those of NETAL in 
the following sense.
BCK's calculation involves one unknown LEC 
(denoted by $L_{\rm 1A}$), 
which like $\hat{d}_R$ in Ref.\cite{PMS-pphep},
represents the strength of  
a four-nucleon axial-current coupling term.
BCK therefore determined $L_{\rm 1A}$
by requiring that
the $\nu d$ cross sections of NETAL be reproduced
by their EFT calculation.
With the value of $L_{\rm 1A}$
fine-tuned this way,
the $\sigma(\nu d)$'s obtained by BCK
show a perfect agreement with
those of NETAL for all the four reactions 
in eq.(\ref{nu-d}) and
for the entire solar neutrino energy range,
$E_\nu\lsim$ 20 MeV.
Moreover, the optimal value,
$L_{\rm 1A}=5.6\,{\rm fm}^3$, found 
by BCK~\cite{EFT} is consistent 
with the order of magnitude of $L_{\rm 1A}$
expected from the naturalness argument 
(based on a dimensional analysis), 
$|L_{\rm 1A}|\le 6\,{\rm fm}^3$.
The fact that an EFT calculation
(with one parameter fine-tuned)
reproduces the results of SNPA very well
strongly suggests the robustness of
the SNPA results for $\sigma(\nu d)$.

Even though it is reassuring 
that the $\nu$-$d$ cross sections
calculated in SNPA and EFT
agree with each other 
(in the sense explained above),
it is desirable to carry out an EFT calculation 
that is free from  any adjustable LEC.
Fortunately, EFT* allows us
to carry out an EFT-controlled parameter-free calculation
of the $\nu$-$d$ cross sections,
and such a calculation was carried out 
by Ando {\it et al.}~\cite{andetal-nud}.
The $\sigma(\nu d)$'s obtained in ~\cite{andetal-nud}
are found to agree within 1\% with
$\sigma(\nu d)$'s obtained by NETAL 
using SNPA~\cite{NETAL}. 
These results show 
that the $\nu$-$d$ cross sections used 
in interpreting the SNO experiments
\cite{ahmetal}
are reliable at the 1\% precision level,
and hence the evidence for neutrino oscillations
reported in those experiments is robust
against nuclear physics ambiguities.

We note that, as PDS~\cite{ksw}
is built on an expansion scheme 
for transition amplitudes themselves,
it does not employ the concept of wave functions.
This feature may be an advantage
in some contexts,
but its disadvantage in the present context
is that one cannot readily relate 
the transition matrix elements
for an A-nucleon system with those for 
the neighboring nuclei;
in PDS, each nuclear system requires a separate 
parametrization.
This feature underlies the fact
that, in the work of BCK~\cite{EFT},
$L_{1A}$ remained undetermined,
as no experimental data is available to fix $L_{1A}$
within the two-nucleon systems.

Although the determination of $\hat{d}_R$
from $\Gamma_{\beta}^t$ should be good enough 
for all practical purposes,
it is worthwhile to study
a possibility to fix $\hat{d}_R$ with the use 
of an observable belonging to the A=2 systems.
A promising candidate is the $\mu$-capture process,
$\mu^-+d\rightarrow \nu_\mu+n+n$.
Although rather large energy-momentum transfers
involved in the disappearance of a $\mu^-$
seem to make the applicability of EFT here a delicate issue,
we can show that, 
as far as the hadron sector is concerned,
$\mu$-$d$ capture is in fact a reasonably ``gentle" process.
This is because: (1) the $\nu_\mu$
carries away most of the energy,
and (2) there is a large enhancement
of the transition amplitude 
in a kinematic region where the relative motion
of the final two nucleons is low enough to
justify the use of EFT.
According to Ando {\it et al.}'s recent
study~\cite{APMK-muD},
$\mu$\,-\,$d$ capture can be useful for
controlling $\hat{d}_R$,
if the quality of experimental data 
improves sufficiently.
We note that an experiment to measure the 
$\mu d$ capture rate with 1\% precision
is planned at the PSI~\cite{kametal}.

\section{Discussion}

In introducing hybrid EFT, we have replaced 
$|\Psi>^{EFT}$ for the initial
and final nuclear states in eq.(\ref{ME-EFT})
with the corresponding $|\Psi>^{SNPA}$'s;
see eq.(\ref{ME-hybrid}).
This replacement may bring in
a certain degree of model dependence,
called the off-shell effect, because
the phenomenological NN interactions 
are constrained only by the 
on-shell two-nucleon observables.\footnote{
In a consistent theory, physical observables
are independent of field transformations
that lead to different off-shell behaviors, 
and therefore the so-called off-shell effect 
is not really a physical effect.
In an approximate theory, observables may 
exhibit superficial dependence on off-shell behavior,
and it is customary to refer to this dependence
as an off-shell effect.} 
This off-shell effect, however, is expected 
to be small for the reactions under consideration,   
since they involve low momentum transfers
and hence are not extremely sensitive to
the short-range behavior of the nuclear wave functions.
One way to quantify this expectation is to compare
a two-nucleon relative wave function generated by the phenomenological
potential with that generated by an EFT-motivated potential.
Phillips and Cohen~\cite{pc00} made such a comparison
in their analysis of the 1-body operators 
responsible for electron-deuteron Compton scattering,
and showed that a hybrid EFT should work well up to
momentum transfer 700 MeV.  
A similar conclusion is expected to hold
for a two-body operator,
so long as its radial behavior
is duly ``smeared-out" reflecting
a finite momentum cutoff. 
Thus, hybrid EFT as applied to low energy
phenomena is expected to be practically
free from the off-shell ambiguities.
The off-shell effect should be even less significant 
in EFT*, wherein an additional ``effective" renormalization
is likely to be at work (see subsection 2.4).
 
I now wish to discuss briefly another very interesting 
development, due to Tom Kuo and his colleagues~\cite{kuo},
which can shed much light
on the reliability of a hybrid EFT or EFT* calculation.
As mentioned, a ``realistic phenomenological" nuclear 
interaction, $V_{ij}$ in eq.(\ref{HSNPA}),
is determined by solving the Schr\"{o}dinger equation,
eq.(\ref{Sch}), for the A=2 system and fitting the results
to the full set of two-nucleon data 
up to the pion production threshold energy. 
So, physically, $V_{ij}$ should reside in a momentum regime
below a certain cutoff, $\Lambda_c$.
In the conventional treatment, however, 
the existence of this cutoff scale is ignored,
and eq.(\ref{Sch}) is solved,
allowing the entire 
momentum range to participate.  
Kuo {\it et al.\ }proposed to construct an 
{\it effective low-momentum} potential 
$V_{low-k}$ by eliminating 
(or integrating out) from $V_{ij}$
the momentum components higher than
$\Lambda_c$, and calculated $V_{low-k}$'s 
corresponding to many well-established examples
of $V_{ij}$'s.  
Remarkably, it was found that all these $V_{low-k}$'s give
identical results for the half-off-shell T-matrices,
even though the ways short-range physics is encoded
in these $V_{ij}$'s are highly diverse.
This implies that the $V_{low-k}$'s are free from
the off-shell ambiguities, and therefore the use of 
$V_{low-k}$'s is essentially equivalent to employing 
$V_{ij}^{EFT}$ (that appeared in eq.(\ref{HEFT})),
which by construction should be model-independent.
Now, as mentioned, our EFT* calculation has
a momentum-cutoff regulator, and this essentially
ensures that the matrix element,
${\cal M}_{fi}^{hyb-EFT}$, in eq.(\ref{ME-hybrid})
is only sensitive to the half-off-shell T-matrices
that are controlled by $V_{low-k}$ instead of $V_{ij}$.
Therefore, we can expect that the EFT* results 
reported here are essentially free from the off-shell
ambiguities.
   
It is also worth noting that 
the calculation of the cross section for ``Hen",
$^3{\rm He}+n\!\rightarrow\!^4{\rm He}+\gamma$,
should provide a further check of the reliability
of EFT*.
Such a calculation is being done by T.-S. Park and 
Y.H. Song~\cite{sp03}.

My final remark in this section is concerned
with the LECs, $L_{1A}$ and $\hat{d}_R$.
Chen, Heeger and Robertson~\cite{chr03}
have recently reported that
a ``self-calibrating" analysis of the SNO data
allows one to place a model-independent
constraint on $L_{1A}$.  
The $\sigma(\nu d)$'s 
corresponding to the range of $L_{1A}$
obtained in this analysis
are consistent with $\sigma(\nu d)$'s obtained 
in SNPA and EFT*.
However, although the method used 
in this self-calibrating analysis is very beautiful,
the resulting constraint on $L_{1A}$
is still rather loose.
From a comparison of the $\sigma(\bar{\nu} d)$'s calculated 
in SNPA and the cross sections measured in a
reactor anti-neutrino experiment~\cite{reactor},
we have known for some time that the theoretical 
values cannot be off by more than 10 \%.
The results of the ``self-calibrating" analysis 
at present do not provide much improvement
over this well-known empirical upper limit of errors. 

I have mentioned that both $L_{1A}$ 
and $\hat{d}_R$ represent the strength
of axial-current-four-nucleon contact coupling.
It is to be noted, however, that 
$L_{1A}$ belongs to pion-less EFT,
while $\hat{d}_R$ to pion-ful EFT.
In the pion-ful EFT, because of the strong tensor force,
the exchange current involving the deuteron $d$-state
is important, and the s-wave exchange current 
arising from the $\hat{d}_R$ term is separate from 
this tensor-force effect.
By contrast, in the pion-less EFT,
the explicit $d$-wave term is a higher-order correction,
and hence the $s$-wave $L_{1A}$ term must subsume
the strong tensor-force contributions.
It would be illuminating to investigate the relation
between  $L_{1A}$ and $\hat{d}_R$ from this perspective.
Such a study is currently underway~\cite{Netal03}

\section{Summary}

After giving a very limited survey
of the current status of nuclear \CPT,
I must repeat my disclaimer
that I have left out many important topics
belonging to nuclear \CPT.
Among others,
I did not discuss very important studies 
by Epelbaum, Gl\"{o}ckle and Mei{\ss}ner\cite{epeetal}
to construct a formally consistent framework
for applying \CPT\ to complex nuclei
(A = 3, 4, ...).  
It should be highly informative
to apply this type of formalism
to electroweak processes and compare 
the results with those of EFT*.
Despite the highly limited scope of topics covered,
I hope I have succeeded in conveying the message  
that EFT* is a reliable framework for 
computing transition amplitudes for a large class
of electroweak processes in light nuclei.
I also wish to emphasize that,
in each of the cases for which both SNPA and EFT* calculations
have been performed, it has been found that 
the result of EFT* supports
and improves the SNPA result.

\vspace{5mm}
\noindent
This talk is based on the work done 
in collaboration with T.-S. Park, M. Rho, 
D.-P. Min, S. Nakamura, T. Sato,
V. Gudkov, F. Myhrer, H. Fearing and Y.H. Song,
and I wish to express my sincere thanks 
to these colleagues.
I also wish to gratefully acknowledge
financial support by the 
US National Science Foundation, 
Grant No. PHY-0140214.

\end{document}